\documentclass[doublecol]{epl2}

\usepackage{graphicx,amssymb}
\usepackage{bm}
\usepackage{amsmath}

\newcommand\beq{\begin{equation}}
\newcommand\eeq{\end{equation}}
\newcommand\beqa{\begin{eqnarray}}
\newcommand\eeqa{\end{eqnarray}}
\newcommand{\dd}{\text{d}}
\newcommand{\ee}{\text{e}}
\newcommand{\nn}{\nonumber\\}

\title{Aging to non-Newtonian hydrodynamics in a granular gas}
\shorttitle{Aging to non-Newtonian hydrodynamics in a granular gas} 

\author{A. Astillero\inst{1} \and A. Santos\inst{2}}
\shortauthor{A. Astillero and A. Santos}  

\institute{
  \inst{1} Departamento de Inform\'atica I, Centro Universitario de
M\'erida, Universidad de Extremadura, E--06800 M\'erida, Spain\\
  \inst{2} Departamento de F\'{\i}sica, Universidad de
Extremadura, E--06071 Badajoz, Spain} 

\abstract{The evolution to the steady state of a granular gas
subject to simple shear flow is analyzed by means of computer
simulations. It is found that, regardless of its initial
preparation, the system reaches (after a transient period lasting a
few collisions per particle) a non-Newtonian (unsteady) hydrodynamic
regime, even at strong dissipation and for states where the time
scale associated with inelastic cooling is shorter than the one
associated with the irreversible fluxes. Comparison with a
simplified rheological model shows a good agreement.}

\begin{document}

\maketitle

\section{Introduction}

A granular gas is a large collection of (mesoscopic or macroscopic)
particles which collide inelastically and are usually kept  in a
state of continuous agitation. Apart from their interest in
industrial and technological applications, granular gases are
important at a fundamental level as physical systems intrinsically
out of equilibrium and thus exhibiting a wide spectrum of complex
behavior \cite{K99,K00,LP00,HM02,JNB96a,BP04}. Although the number
of grains in a fluidized granular system is of course much smaller
than the number of atoms or molecules in a conventional gas, it is
large enough as to make nonequilibrium statistical-mechanical
concepts and tools applicable. In particular, a kinetic theory
approach (based on the Boltzmann and Enskog equations suitably
modified to account for inelastic collisions) has proven to be very
useful \cite{BP04}. However, because of the energy dissipation upon
collisions and the associated lack of detailed balance and Gibbs
equilibrium state, one is not allowed to take for granted any
phenomenology that applies to normal gases, unless much caution is
exercised. Quoting Kadanoff, ``one might even say that the study of
granular materials gives one a chance to reinvent statistical
mechanics in a new context'' \cite{K99}.

Hydrodynamics is one of the key features of standard fluids. As is
well known, the hydrodynamic description of a conventional fluid
consists of closing the exact balance equations for the densities of
the conserved quantities (mass, momentum and energy) with
constitutive equations relating the momentum and heat fluxes to the
conserved densities (usually referred to as hydrodynamic fields) and
their gradients. If the hydrodynamic gradients are weak, the fluxes
can be assumed to be linear in those gradients, what results in the
Navier--Stokes (NS) hydrodynamic description. On the other hand,
even if the gradients are strong, a (non-Newtonian) hydrodynamic
regime beyond the NS one is still possible. The conventional
scenario for the ``aging to hydrodynamics'' in a normal gas can be
summarized as follows \cite{DvB77}. Given an arbitrary initial
state, the evolution proceeds along two successive stages. First,
during the so-called \emph{kinetic} stage there is a fast relaxation
(lasting a few collision times) to a ``universal'' (or ``normal'')
velocity distribution $f(\mathbf{r},\mathbf{v};t)$ that is a
functional of the hydrodynamic fields (number density, flow velocity
and temperature). Subsequently, the \emph{hydrodynamic} stage is
described through a slower evolution of the hydrodynamic fields as
they approach equilibrium or an externally imposed nonequilibrium
steady state. The first stage is sensitive to the initial
preparation of the system, while in the hydrodynamic regime the
system has practically ``forgotten'' the details of its initial
state (except for an implicit dependence on the initial conditions
through the hydrodynamic fields). If the hydrodynamic gradients are
small enough when the hydrodynamic state is reached, the latter can
be described by the NS terms in the Chapman--Enskog expansion
\cite{CC70}. However, a normal (or hydrodynamic) velocity
distribution function is not restricted to the NS domain but can
apply to the non-Newtonian regime as well \cite{GS03}.

The applicability of a hydrodynamic description to granular fluids
is not self-evident at all \cite{K99}. In particular, the absence of
energy conservation gives rise to a sink term in the energy balance
equation which might preclude, except perhaps in quasi-elastic
situations, the role of the (granular) temperature as a hydrodynamic
variable. At least in the NS regime, however, there exists
compelling evidence from theory \cite{BD05}, simulations \cite{sim}
and experiments \cite{exp} supporting the validity of a hydrodynamic
treatment of granular fluids. On the other hand, the inherent lack
of scale separation in sheared granular gases invalidates a NS
description, so that the applicability of hydrodynamics in those
cases has been controversial \cite{TG98}.

The scenario of aging to hydrodynamics described above is known to
hold in two limiting cases: finite dissipation but no gradients
(\emph{i.e.}, in the approach to the homogeneous cooling state)
\cite{DHGD02} and zero dissipation but large gradients \cite{GS03}.
Having that in mind, the question we want to address in this paper
is, does that scenario still apply to granular gases with
\emph{strong} dissipation in the \emph{non-Newtonian} regime?

\section{Uniform shear flow}
In order to contribute to an understanding of the previous question
and its possible answer, we will  focus on inelastic hard spheres in
the so-called simple or uniform shear flow (USF), which is
characterized by a constant density $n$, a time-dependent (but
uniform) granular temperature $T(t)$ and a flow velocity with a
linear profile $\mathbf{u}(\mathbf{r})=a y \widehat{\mathbf{x}}$,
$a$ being the constant shear rate. There are three basic reasons to
choose this state to isolate the problem at hand. First, it is
macroscopically simple since only a hydrodynamic gradient exists
($a=\partial u_x/\partial y$) and moreover it is a constant. Second,
this flow possesses a steady state (resulting from the balance
between inelastic cooling and viscous heating) that is inherently
non-Newtonian \cite{TG98,SGD04}. Finally, since both the density $n$
and the flow velocity $\mathbf{u}(\mathbf{r})$ are independent of
time, the possible aging to hydrodynamics is enslaved by the
temporal evolution of the granular temperature, which is precisely
the quantity casting doubts on the applicability of hydrodynamics.

In the USF the mass continuity equation is identically satisfied,
while  momentum conservation implies that the shear stress $P_{xy}$
is uniform. The energy balance equation becomes
\begin{equation}
\partial _{t}T(t)=-({2}a/{3n})P_{xy}(t) -\zeta(t) T(t),
\label{2.8}
\end{equation}
where $\zeta(t)$ is the cooling rate due to inelasticity. Given a
shear rate $a$ and a coefficient of normal restitution $\alpha$, a
stationary temperature $T_s=\lim_{t\to\infty} T(t)$ is reached when
the viscous heating and the inelastic cooling cancel each other.
This steady state can be conveniently described by introducing
dimensionless quantities. Let $\nu(T)=nT/\eta_0(T)\propto n
T^{1/2}$, where $\eta_0(T)$ is the NS shear viscosity in the elastic
case,  denote an effective collision frequency. We can then define a
reduced shear rate $a^*(t)=a/\nu(T(t))$, a reduced cooling rate
$\zeta^*(t)=\zeta(t)/\nu(T(t))$ and a reduced shear stress
$P_{xy}^*(t)=P_{xy}(t)/nT(t)$. Thus, one has
$\frac{2}{3}a_s^*|P_{xy,s}^*|=\zeta_s^*$ in the steady state. The
stationary temperature is approached from below ($T(t)<T_s$) if the
shear rate is large enough and/or the inelasticity is small enough
so that the heating rate (due to viscous effects)
$\frac{2}{3}a^*(t)|P_{xy}^*(t)|$ prevails over the cooling rate (due
to inelasticity) $\zeta^*(t)$; we will refer to this situation as
the \emph{heating} case. Otherwise,  the stationary temperature is
reached from above  and this will be referred to as the
\emph{cooling} case. In this context, the question posed above can
be rephrased as, does an \emph{unsteady} hydrodynamic regime exist
in both the heating and cooling cases
 \emph{before} the steady state is achieved?

Restricting ourselves to low density granular gases and to states
that are uniform in the Lagrangian frame of reference, so that
$f(\mathbf{r},\mathbf{v};t)=f(\mathbf{V}(\mathbf{r}),t)$ with the
peculiar velocity
$\mathbf{V}(\mathbf{r})=\mathbf{v}-\mathbf{u}(\mathbf{r})$, the
Boltzmann equation becomes \cite{GS03,SGD04}
\begin{equation}
\left( \partial _{t}-aV_{y}{\partial_{V_{x}}}\right)
f(\mathbf{V},t)=J[\mathbf{V}|f(t)],  \label{a.11}
\end{equation}
where $J[\mathbf{V}|f(t)]$ is the (inelastic) Boltzmann collision
operator. Equation \eqref{a.11} must be complemented with an initial
condition $f(\mathbf{V},0)=f^0(\mathbf{V})$, so that the solution is
actually a functional of the initial distribution, \emph{i.e.},
$f(\mathbf{V},t)=f(\mathbf{V},t|f^0)$. Analogously, the velocity
moments of $f$, such as the pressure tensor $P_{ij}(t|f^0)$ are also
functionals of $f^0$. Since the only time-dependent hydrodynamic
variable is the temperature $T(t)$ and the only hydrodynamic
gradient is the shear rate $a$, the existence of a hydrodynamic
regime implies that, after a certain number of collisions per
particle,
\beq
f(\mathbf{V},t|f^0)\to n\left[{m}/{2 T(t)}\right]^{3/2}
f^*(\mathbf{C}(t);a^*(t)),
\label{1}
\eeq
where $\mathbf{C}(t)\equiv \mathbf{V}/\sqrt{2T(t)/m}$ is the
(peculiar) velocity in units of the (time-dependent) thermal speed,
$m$ being the mass of a grain. The scaled velocity distribution
function $f^*(\mathbf{C};a^*)$ is, for a given value of the
coefficient of restitution $\alpha$, a \emph{universal} function in
the sense that it is independent of the initial state $f^0$ and
depends on the applied shear rate $a$ through the reduced scaled
quantity $a^*$ only. In other words, if a hydrodynamic description
is possible, the different solutions $f(\mathbf{V},t|f^0)$ of the
Boltzmann equation \eqref{a.11} would be ``attracted'' by the
universal form \eqref{1}. For asymptotically long times, the steady
state would eventually be reached, \emph{i.e.},
$f^*(\mathbf{C};a^*)\to f_s^*(\mathbf{C})=f^*(\mathbf{C};a_s^*)$.
Equation \eqref{1} has its counterpart at the level of the velocity
moments. In particular, the pressure tensor $P_{ij}(t|f^0)$ would
become
\beq
P_{ij}(t|f^0)\to nT(t)P_{ij}^*(a^*(t))
\label{2}
\eeq
with universal functions $P_{ij}^*(a^*)$. The non-Newtonian
character of the hydrodynamic regime can be characterized by the
nonlinear (reduced) shear viscosity $\eta^*(a^*)$ and viscometric
function $\Psi^*(a^*)$ defined by
\beq
\eta^*(a^*)=-\frac{P_{xy}^*(a^*)}{a^*}, \quad
\Psi^*(a^*)=\frac{P_{xx}^*(a^*)-P_{yy}^*(a^*)}{{a^*}^2}.
\eeq

\section{Rheological model}
Before presenting the simulation results, it is worth considering a
simple kinetic model \cite{BDS99} in which the Boltzmann collision
operator $J[\mathbf{V}|f(t)]$ is replaced by a relaxation-time term
toward the local equilibrium distribution plus a drag-force term
mimicking the cooling effects due to inelastic collisions. Applied
to the USF, it closes eq.\ \eqref{2.8} with the evolution equations
\cite{SGD04,BRM97,SA05}:
\beq
\partial_t P_{xy}=-a
P_{yy}-\beta\nu(T)P_{xy}-\zeta(T)P_{xy},
\label{5}
\eeq
\beq
\partial_t P_{yy}=-\beta\nu(T)\left(P_{yy}-nT\right)-\zeta(T)P_{yy},
\label{6}
\eeq
where  $\zeta(T)=\zeta^*\nu(T)$ with
$\zeta^*=\frac{5}{12}(1-\alpha^2)$ and here we take
$\beta=\frac{1}{2}(1+\alpha)$. Equations \eqref{5} and \eqref{6} can
also be obtained, with $\beta=\frac{1}{6}(1+\alpha)(2+\alpha)$, from
the Boltzmann equation in Grad's approximation \cite{SGD04,G02}.

The set of coupled equations \eqref{2.8}, \eqref{5} and \eqref{6}
must be solved numerically \cite{SGD04}, the solution including both
the  transient kinetic regime and the hydrodynamic evolution stage.
In order to extract the hydrodynamic solution  in an
\emph{analytical} way, we will introduce an additional
simplification similar to that carried out in the elastic case
\cite{GS03,KDN97}. As mentioned before, $\nu(T)\propto n T^q$ with
$q=\frac{1}{2}$. However, here we will temporarily view $q$ as a
(small) free parameter, so that the solutions to  eqs.\ \eqref{2.8},
\eqref{5} and \eqref{6} depend parametrically on $q$ \cite{ETB06}.
If $q=0$, then $\nu=\text{const}$ and $a^*=\text{const}$, so there
is no steady state (except if $a^*$ takes a specific value).
However, for sufficiently long times the scaled quantities
$P_{ij}^*$ reach well defined (hydrodynamic) values which depend on
$a^*$ and not on the initial state \cite{AS07}. Next, carrying out a
first-order perturbation analysis around $q=0$ and constructing
Pad\'e approximants, one finally gets \cite{AS07}
\beqa
\eta^*(a^*)&=&\beta^{-1}
\left[1+2\gamma(a^*)\right]^{-2}\Bigl\{1+\frac{1}{2}
\left[\zeta^*/\beta-2\gamma(a^*)\right]\nn
&&\times\left[{1-6\gamma(a^*)}\right]/{\left[1+6\gamma(a^*)\right]^2}\Bigr\}^{-1},
\label{8}
\eeqa
\beqa
\Psi^*(a^*)&=&2 \beta^{-2}
\left[1+2\gamma(a^*)\right]^{-3}\Bigl\{1+\frac{
\zeta^*/\beta-2\gamma(a^*)}{2\left[1+6\gamma(a^*)\right]^2}\Bigr\}^{-1}\nn
&&\times \Bigl\{1+\frac{
\zeta^*/\beta-2\gamma(a^*)}{\left[1+6\gamma(a^*)\right]^2}\Bigr\}^{-1},
\label{9}
\eeqa
where $\gamma(a^*)\equiv
\frac{2}{3}\sinh^2\left[\frac{1}{6}\cosh^{-1}(1+9{a^*}^2/\beta^2)\right]$
and we have already set $q=\frac{1}{2}$. Equations \eqref{8} and
\eqref{9} provide the hydrodynamic non-Newtonian  viscosity and
viscometric function as  explicit functions of the reduced shear
rate $a^*$ (and of the coefficient of restitution $\alpha$) in our
simplified rheological model. The steady state corresponds to just
one point on the curves $\eta^*(a^*)$ and $\Psi^*(a^*)$. In the
model, it is given by $a^*_s=\sqrt{3\zeta^*/2\beta}(\beta+\zeta^*)$
\cite{SGD04,BRM97,SA05}, which verifies the condition
$\gamma(a_s^*)=\zeta^*/2\beta$, and so
$\eta_s^*=\eta^*(a_s^*)=\beta/(\beta+\zeta^*)^2$ and
$\Psi_{s}^*=\Psi^*(a_s^*)=2\beta/(\beta+\zeta^*)^3$. If we formally
take the limit of vanishing shear rate, we recover the NS viscosity
of the inelastic gas predicted by the model \cite{SA05}, namely
$\lim_{a^*\to 0}\eta^*(a^*)=(\beta+\zeta^*/2)^{-1}$, as well as the
Burnett value $\lim_{a^*\to
0}\Psi^*(a^*)=2/(\beta+\zeta^*/2)(\beta+\zeta^*)$. We have checked
that, for $\alpha\geq 0.5$, Eqs.\ \eqref{8} and \eqref{9} yield
values which are hardly distinguishable from those obtained by a
numerical solution of the set \eqref{2.8}, \eqref{5} and \eqref{6}
\cite{AS07}.

\section{Monte Carlo simulations}
In order to test whether the hydrodynamic regime \eqref{1} exists or
not, we have solved the Boltzmann equation \eqref{a.11} by means of
the direct simulation Monte Carlo (DSMC) method  \cite{B94}. As in
ref.\ \cite{AS05}, we have taken $N=10^4$ simulated particles and a
varying time step $\delta t=10^{-3}\tau^0\sqrt{T^0/{T}}$, where
$T^0$ is the initial temperature and
$\tau^0=\lambda/\sqrt{2T^0/m}\simeq 0.8885\nu^{-1}(T^0)$ is the
initial mean free time, $\lambda$ being the mean free path. In order
to improve the statistics, the results are averaged over 100
independent realizations. The values of the coefficient of
restitution considered have been $\alpha=0.5$, $0.7$ and $0.9$. For
each value of $\alpha$, four different shear rates have been taken:
$a=0.01/\tau^0$, $a=0.1/\tau^0$, $a=4/\tau^0$ and $a=10/\tau^0$. The
two first values of the shear rate ($a=0.01/\tau^0$ and
$a=0.1/\tau^0$) are small enough to correspond to cooling cases,
even for the least inelastic system ($\alpha=0.9$), while the other
two values ($a=4/\tau^0$ and $a=10/\tau^0$) are large enough to
correspond to heating cases, even for the most inelastic system
($\alpha=0.5$). For each one of the twelve pairs $(\alpha,a)$, five
widely different initial conditions have been chosen, so that sixty
independent states have been simulated. The first initial condition
(here referred to as A) is just the (local) equilibrium one, namely
\beq
f^0(\mathbf{V})=n\left({m}/{2\pi T^0}\right)^{3/2}\ee^{-mV^2/2T^0}.
\label{3}
\eeq
The other four initial conditions are of the anisotropic form
\begin{eqnarray}
f^0(\mathbf{V})&=&({n}/{2})\left({m}/{2\pi T^0}\right)^{1/2}
\ee^{-{m V_{z}^{2}}/{2T^0}}\nn
&&\times\left[\delta\left(V_{x}-V^{0}\cos
\phi\right)\delta\left(V_{y}+V^{0}
\sin \phi\right)\right. \nonumber \\
&&\left.+\delta\left(V_{x}+V^{0}\cos
\phi\right)\delta\left(V_{y}-V^{0}\sin \phi\right)\right],
\label{4}
\end{eqnarray}
where $\delta(x)$ is Dirac's distribution, $V^0\equiv\sqrt{2T^0/m}$
is the initial thermal speed and $\phi\in [0,\pi]$ is an  angle
characterizing each specific condition.   The pressure tensor
corresponding to eq.\ \eqref{4} is given by
$P_{xx}^0=2nT^0\cos^2\phi$, $P_{yy}^0=2nT^0\sin^2\phi$,
$P_{zz}^0=nT^0$ and $P_{xy}^0=-nT^0\sin 2\phi$. The four values of
$\phi$ considered are $\phi=k \pi/4$ with $k=0$, $1$, $2$ and $3$;
we will denote the respective initial conditions of type \eqref{4}
as B0, B1, B2 and B3. Note that $P_{xy}^0=0$ in the initial
conditions A, B0 and B2, while $P_{xy}^0=-nT^0$ in the case of B1.
On the other hand, $P_{xy}^0$ takes an ``artificial'' positive value
($P_{xy}^0=nT^0$) if the system is initially prepared with the
condition B3. According to eq.\ \eqref{2.8}, this implies that
during the very first times the fluid undergoes a double cooling
effect: the inelastic cooling plus an artificial shear stress
cooling. It takes some time before a ``natural'' negative value of
$P_{xy}(t)$ is established  and the usual viscous heating effect
competes against the inelastic cooling.

\section{Results and discussion}
\begin{figure}
\includegraphics[width=1 \columnwidth]{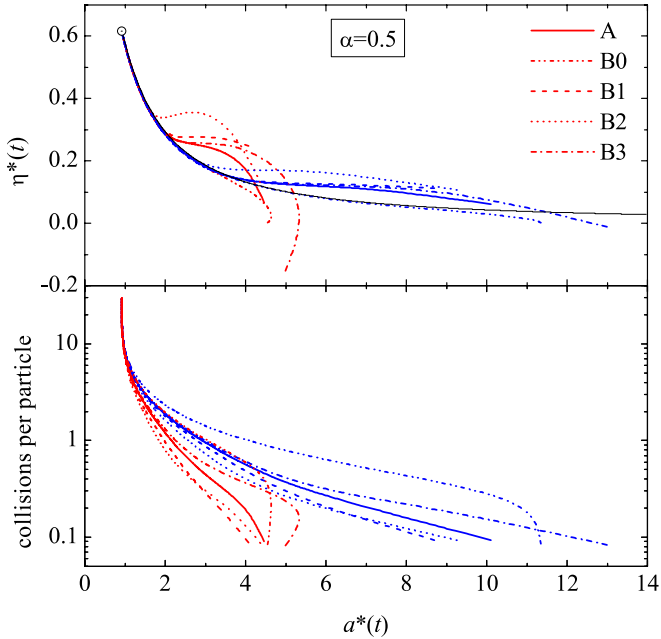}
\caption{(Color online) Reduced shear viscosity $\eta^*(t)$ (top
panel) and number of collisions per particle (bottom panel) versus
the reduced shear rate $a^*(t)$ for $\alpha=0.5$ in the two heating
cases $a=4/\tau^0$ (red lines) and $a=10/\tau^0$ (blue lines). In
the top panel, the circle represents the steady-state point
$(a_s^*,\eta_s^*)$, while the thin solid line corresponds to the
hydrodynamic function, eq.\ \protect\eqref{8}, obtained from our
simplified rheological model. Note the logarithmic scale in the
vertical axis of the bottom panel.}
\label{fig1}
\end{figure}
The solution of the Boltzmann equation \eqref{a.11} through the DSMC
method allows one to monitor the temporal evolution of velocity
moments, such as the temperature $T(t)$ and the pressure tensor
$P_{ij}(t)$, as well as of the velocity distribution function
$f(\mathbf{V},t)$ itself. In particular, one can follow the reduced
shear rate $a^*(t)=a/\nu(T(t))$ (which decreases in the heating
states and increases in the cooling states) and the reduced shear
viscosity $\eta^*(t)=-P_{xy}(t)/nT(t)a^*(t)$. A parametric plot of
$\eta^*(t)$ vs $a^*(t)$ is then useful to test the establishment of
a hydrodynamic regime in which $\eta^*(t)\to \eta^*(a^*(t))$, where
the function $\eta^*(a^*)$ must be independent of the initial
conditions. Such a parametric plot is shown in the top panel of
fig.\ \ref{fig1} for the most inelastic gas ($\alpha=0.5$) and for
the two heating cases ($a=4/\tau^0$ and $a=10/\tau^0$). The curve
representing the rheological model \eqref{8} is also included. It is
quite apparent that the ten curves are attracted to a common
universal curve which, in addition, turns out to be excellently
described by our simplified model. The bottom panel of fig.\
\ref{fig1} displays the temporal evolution  of $a^*(t)$, as measured
by the accumulated number of collisions per particle (\emph{i.e}.,
the total number of collisions in the simulations divided by the
total number of particles). We can observe that the states with the
highest heating effect ($a=10/\tau^0$) reach the hydrodynamic stage
after about 1 collision per particle only, while this aging period
takes a little longer (about 2 collisions per particle) in the
states with $a=4/\tau^0$. Once the state joins the hydrodynamic
curve, it moves along it until the steady state is reached, what
occurs after typically 10 collisions per particle since the initial
state. This means that about 80--90\% of the duration (measured by
the number of collisions) of the evolution to the steady state is
occupied by the hydrodynamic stage. It is also interesting to remark
that, for a given value of $a$, the slowest heating takes place for
the initial condition B0, followed by B3, A and B1, while the
fastest heating corresponds to B2. This implies that the state
starting from B0 is the one joining the hydrodynamic curve at a
larger value of $a^*$ ($a^*\simeq 7$ and $a^*\simeq 3$ for
$a=10/\tau^0$ and $a=4/\tau^0$, respectively), while the one
starting from B2 does it at a smaller value ($a^*\simeq 3$ and
$a^*\simeq 1.5$ for $a=10/\tau^0$ and $a=4/\tau^0$, respectively).

\begin{figure}
\includegraphics[width=1 \columnwidth]{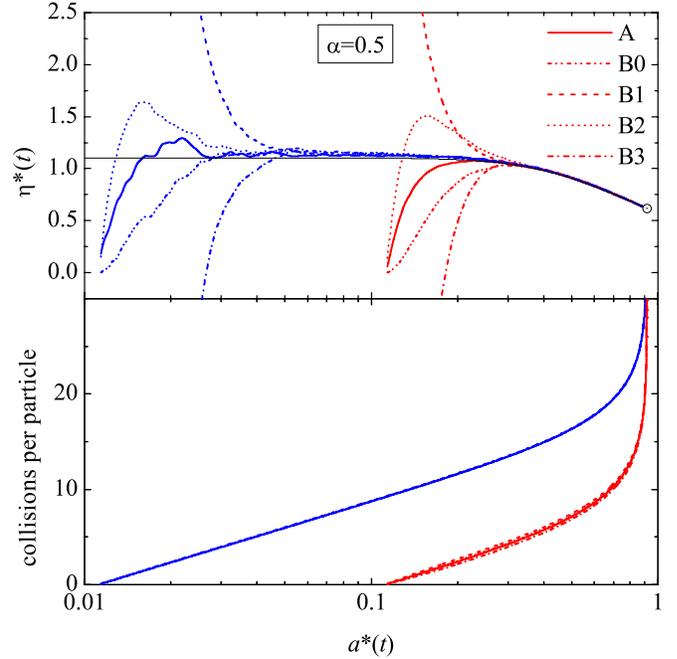}
\caption{(Color online) Same as in fig.\ \protect\ref{fig1}, but for
 the two cooling cases $a=0.01/\tau^0$ (blue lines) and
$a=0.1/\tau^0$ (red lines). Note the logarithmic scale in the
horizontal axes and the linear scale in the vertical axis of the
bottom panel.}
\label{fig2}
\end{figure}
Let us consider now the  cases $a=0.01/\tau^0$ and $a=0.1/\tau^0$.
The temporal evolution is dominated now by the inelastic cooling
(\emph{i.e.}, the time scale $\zeta^{-1}$ associated with the
cooling is shorter than the time scale $\frac{3}{2} nT/a|P_{xy}|$
associated with the viscous heating) and so the existence of a
hydrodynamic regime might appear as more doubtful than in the
heating cases. That this is not actually the case can be concluded
from fig.\ \ref{fig2}. Again, the ten curves tend to collapse to a
common curve and again the latter practically coincides with the
theoretical prediction \eqref{8}, except perhaps near the NS region
of small $a^*$. In contrast to the heating cases depicted in fig.\
\ref{fig1}, however, the temporal evolution of $T(t)$, and hence of
$a^*(t)\propto T^{-1/2}(t)$, is practically independent of the
initial condition, especially in the case $a=0.01/\tau^0$. This is a
consequence of the fact that for these low values of $a\tau^0$ the
viscous term in eq.\ \eqref{2.8} can be neglected for short times
and so the temperature initially evolves as in the homogeneous
cooling state (decaying practically exponentially with the number of
collisions), hardly affected by the details of the initial state. By
the time the viscous heating effect becomes comparable to the
inelastic cooling, the hydrodynamic regime has already been
attained. It is important to note that the number of collisions
needed to reach the hydrodynamic behavior (about 5 and 10 collisions
per particle for $a=0.01/\tau^0$ and $a=0.1/\tau^0$, respectively)
is longer in the cooling cases than in the heating cases. On the
other hand, the total duration of the evolution period is also
longer than in the heating cases and so the granular gas still lies
on the hydrodynamic curve for  most of the evolution time (as
measured by the number of collisions per particle). The duration of
the transient kinetic stage  in the cooling cases (about 5--10
collisions per particle) is consistent with the one observed in
molecular dynamics simulations of the homogeneous cooling state
\cite{DHGD02}.

We have observed  behaviors similar to those of figs.\ \ref{fig1}
and \ref{fig2} for the viscometric function and for the other
inelasticities ($\alpha=0.7$ and $\alpha=0.9$). While the number of
collisions the system needs to loss memory of its initial state is
practically independent of $\alpha$, the total duration of the
evolution period tends to increase with $\alpha$, especially in the
cooling cases \cite{AS07}. This means that the more elastic the
system, the smaller the fraction of time spent in the kinetic regime
prior to the hydrodynamic stage.

\begin{figure}
\includegraphics[width=1 \columnwidth]{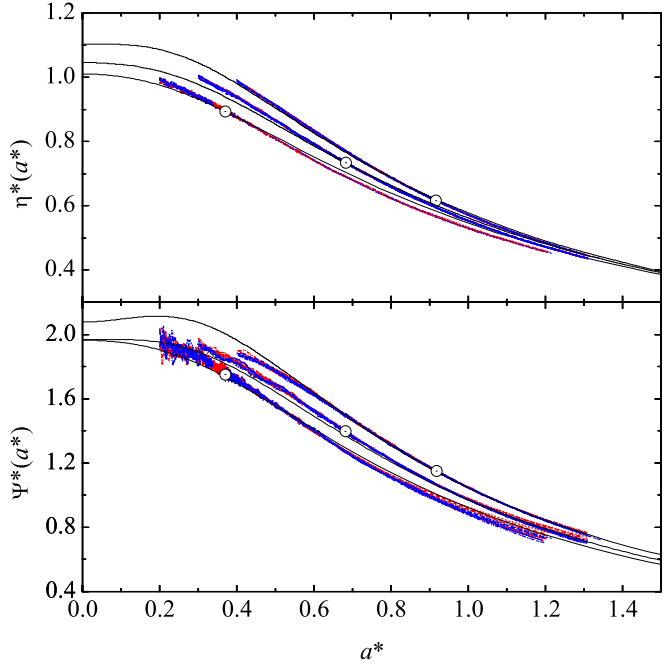}
\caption{(Color online) Reduced shear viscosity $\eta^*(a^*)$ (top
panel) and  viscometric function $\Psi^*(a^*)$ (bottom panel) for
the hydrodynamic part of the evolution to the steady state
(represented by a circle) for, from top to bottom, $\alpha=0.5$, 0.7
and 0.9. The color code is the same as in figs.\ \protect\ref{fig1}
and \protect\ref{fig2}. The thin solid lines correspond to the
hydrodynamic functions, eqs.\ \protect\eqref{8} and
\protect\eqref{9}, obtained from our simplified rheological model.}
\label{fig3}
\end{figure}
The common hydrodynamic parts of the functions $\eta^*(a^*)$ and
$\Psi^*(a^*)$ for the ten heating cases and the ten cooling cases
are displayed in fig.\ \ref{fig3} for each one of the three
inelasticities considered. This corresponds to the shear rate
windows $0.4\leq a^*\leq 1.3$, $0.3\leq a^*\leq 1.3$ and $0.2\leq
a^*\leq 1.2$ for $\alpha=0.5$, 0.7 and 0.9, respectively. The degree
of overlapping of the curves is excellent, although the statistical
fluctuations inherent to the DSMC method are more important for
$\Psi^*(a^*)$ than for $\eta^*(a^*)$, especially in the cooling
branch of the curves. It is also apparent that our simplified model,
eqs.\ \eqref{8} and \eqref{9}, describes reasonably well the
rheological properties of the sheared granular gas

\begin{figure}
\includegraphics[width=1 \columnwidth]{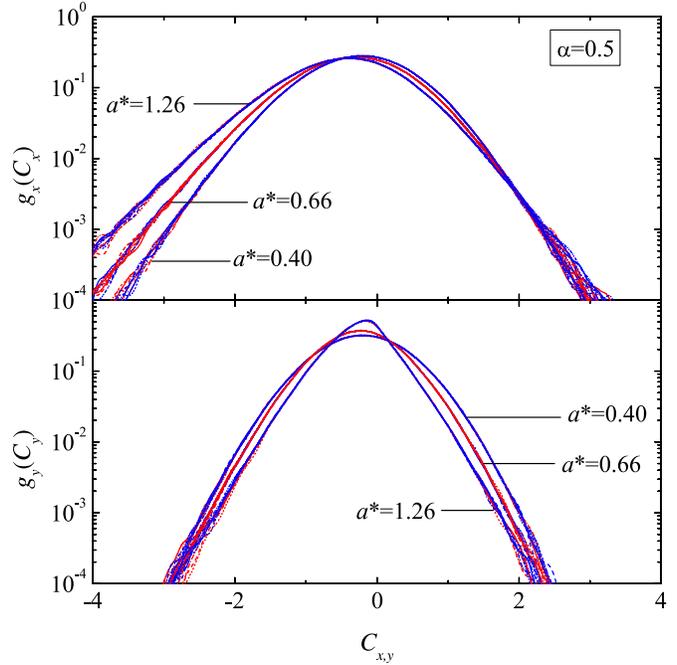}
\caption{(Color online) Marginal distribution functions
$g_{x}(C_{x})$ and $g_{y}(C_{y})$ for $\alpha=0.5$ and $a^*=0.40$,
0.66 and 1.26. The color code is the same as in figs.\
\protect\ref{fig1} and \protect\ref{fig2}.}
\label{fig4}
\end{figure}
Figures \ref{fig1}--\ref{fig3} confirm eq.\ \eqref{2}, \emph{i.e.},
the existence of well defined hydrodynamic rheological functions
$P_{ij}^*(a^*)$ acting as attractors  in the evolution of the
pressure tensor $P_{ij}(t|f^0)$, regardless of the initial
preparation $f^0$. This in turn provides indirect support to the
stronger statement \eqref{1}. To test it in a more direct way, we
have considered the marginal distributions \cite{AS05}
$g_{x}(C_{x};a^*)$ and $g_{y}(C_{y};a^*)$, defined by
\beq
g_{x,y}(C_{x,y};a^*)=\int_{-\infty}^\infty \dd C_z\int_0^\infty \dd
C_{y,x} f^*(\mathbf{C};a^*),
\label{10}
\eeq
for $\alpha=0.5$ and three values of $a^*$ that, according to figs.\
\ref{fig1}--\ref{fig3}, are within the hydrodynamic range, namely
$a^*=0.40$, 0.66 (cooling branch) and 1.26 (heating branch). The
results are plotted in fig.\ \ref{fig4}, which shows an excellent
overlapping of the ten curves for each value of $a^*$, although
obviously the tails of the distribution exhibit larger statistical
fluctuations than the thermal region. As expected, the anisotropic
features of the velocity distribution increase with the shear rate.
The steady state corresponds to $a_s^*=0.92$ \cite{AS05} and hence
the shapes of the associated steady distributions (not shown) lie in
between those for $a^*=0.66$ and $a^*=1.26$.

In conclusion, the results reported in this paper, along with a more
extensive study that will be published elsewhere \cite{AS07},
strongly support that the conventional scenario of aging to
hydrodynamics remains essentially valid for granular gases, even at
high dissipation, for non-Newtonian states  and for situations where
the time scale associated with inelastic cooling is shorter than the
one associated with the irreversible fluxes.

\acknowledgments

Support from the Ministerio de Educaci\'on y Ciencia (Spain) through
grant No.\ FIS2004-01399 (partially financed by FEDER funds) is
gratefully acknowledged.

\end{document}